\documentclass[twocolumn,prl,showpacs,preprintnumbers,superscriptaddress]{revtex4}
\usepackage{graphicx}

\def\gta{\mathrel{\raise 0.1em \hbox{$>$} \hskip -0.8em \lower 0.4em
   \hbox{$\sim$}}}

\begin{document}

\preprint{MIT-CTP-3183}

\title{Inflationary spacetimes are not past-complete}

\author{Arvind Borde}
 \affiliation{Institute of Cosmology, Department of Physics and Astronomy\\
     Tufts University, Medford, MA 02155, USA.}
 \affiliation{Natural Sciences Division, Southampton College, NY~11968, USA.}
\author{Alan H. Guth}
 \affiliation{Institute of Cosmology, Department of Physics and Astronomy\\
     Tufts University, Medford, MA 02155, USA.}
 \affiliation{Center for Theoretical Physics,
     Laboratory for Nuclear Science and Department of Physics,
     Massachusetts Institute of Technology, Cambridge,
     MA 02139, USA.}
\author{Alexander Vilenkin}
 \affiliation{Institute of Cosmology, Department of Physics and Astronomy\\
     Tufts University, Medford, MA 02155, USA.}

\date{January 11, 2003}

\begin{abstract}
Many inflating spacetimes are likely to violate the weak energy
condition, a key assumption of singularity theorems. Here we
offer a simple kinematical argument, requiring no energy
condition, that a cosmological model which is inflating~-- or
just expanding sufficiently fast~-- must be incomplete in null
and timelike past directions.  Specifically, we obtain a bound on
the integral of the Hubble parameter over a past-directed
timelike or null geodesic.  Thus inflationary models require
physics other than inflation to describe the past boundary of the
inflating region of spacetime.
\end{abstract}


\maketitle

\noindent \textit{I. Introduction.}
Inflationary cosmological models~\cite{GuthInf, LindeInf,
AlbStein} are generically eternal to the
future~\cite{VilenkinEt,LindeOne}. In these models, the Universe
consists of post-inflationary, thermalized regions coexisting
with still-inflating ones. In comoving coordinates the
thermalized regions grow in time and are joined by new
thermalized regions, so the \textit{comoving volume} of the
inflating regions vanishes as $t\to\infty$. Nonetheless, the
inflating regions expand so fast that their
\textit{physical volume} grows exponentially with time. As a
result, there is never a time when the Universe is completely
thermalized. In such spacetimes, it is natural to ask if the
Universe could also be past-eternal. If it could, eternal
inflation would provide a viable model of the Universe with no
initial singularity.  The Universe would never come into
existence. It would simply exist.

This possibility was discussed in the early days of inflation,
but it was soon realized~\cite{ST,LI} that the idea could not be
implemented in the simplest model in which the inflating universe
is described by an exact de Sitter space.  More general theorems
showing that inflationary spacetimes are geodesically incomplete
to the past were then proved~\cite{BVOne}.
One of the key assumptions made in these theorems is that the
energy-momentum tensor obeys the weak energy condition.  Although
this condition is satisfied by all known forms of classical
matter, subsequent work has shown that it is likely to be
violated by quantum effects in inflationary
models~\cite{BVFour,GuVaWi}.  Such violations must occur whenever
quantum fluctuations result in an increase of the Hubble
parameter $H$ --- i.e., when $dH/dt > 0$ --- provided that the
spacetime and the fluctuation can be approximated as locally
flat.  Such upward fluctuations in $H$ are essential for the
future-eternal nature of chaotic inflation.  Thus, the weak
energy condition is generally violated in an eternally inflating
universe.
These violations
appear to open the door again to the possibility that inflation,
by itself, can eliminate the need for an initial singularity.
Here we argue that this is not the case. In fact, we show that
the general situation is very similar to that in de Sitter space.

The intuitive reason why de Sitter inflation cannot be
past-eternal is that, in the full de Sitter space, exponential
expansion is preceded by exponential contraction.  Such a
contracting phase is not part of standard inflationary models,
and does not appear to be consistent with the physics of
inflation.  If thermalized regions were able to form all the way
to past infinity in the contracting spacetime, the whole universe
would have been thermalized before inflationary expansion could
begin.  In our analysis we will exclude the possibility of such a
contracting phase by considering spacetimes for which the past
region obeys an {\em averaged expansion condition}, by which we
mean that the average expansion rate in the past is greater than
zero:
\begin{equation}
H_{\rm av} > 0.
\label{Hbound}
\end{equation}
With a suitable definition of $H$ and the region over which the
average is to be taken, we will show that the averaged expansion
condition implies past-incompleteness.

It is important to realize that the terms expansion and
contraction refer to the behavior of congruences of timelike
geodesics (the potential trajectories of test particles).  It is
meaningless to say that a spacetime is expanding at a single
point, since in the vicinity of any point one can always
construct congruences that expand or contract at any desired
rate.  We will see, however, that nontrivial consequences can
result if we assume the existence of a single congruence with a
positive average expansion rate throughout some specified region.

While the past of an inflationary model is a matter of
speculation, the attractor nature of the inflationary equations
implies that many properties of the future can be deduced
unambiguously. According to the standard picture of inflation, all
physical quantities are slowly varying on the scale of $H^{-1}$.
In the vicinity of any point $P$ in the inflating region, we can choose
an approximately homogenous, isotropic and flat spacelike surface
which can serve as the starting point for the standard analysis
of stochastic evolution~\cite{GLM}.
A simple pattern of expansion is established, in which the comoving
geodesics ${\bf x}=\hbox{\it const}$ in the synchronous gauge
form a congruence with $H \gta  \sqrt{(8 \pi /3) G \rho_0}$,
where $\rho_0$ is the minimum energy
density of the inflationary part of the potential energy
function.  This congruence covers the future light cone
of~$P$. While large fluctuations can drive $H$ to negative
values, such fluctuations are extremely rare. Once inflation ends
in any given region, however, many of the geodesics are likely to
develop caustics as the matter clumps to form galaxies and black
holes.
If we try to describe inflation that is
eternal into the past, it would seem reasonable to assume that the past
of~$P$ is like the inflating region to
the future, which would mean that a congruence that is expanding
everywhere, except for rare fluctuations,
can be defined throughout that past.

For the proof of our theorem, however, we find that it is
sufficient to adopt a much weaker assumption, requiring only that
a congruence with $H_{\rm av} > 0$ can be continuously defined
along some past-directed timelike or null geodesic.

In Section~II, we illustrate our result by showing how it arises
in the case of a homogeneous, isotropic, and spatially flat
universe. In the course of the argument we shed some light on the
meaning of an incomplete null geodesic by relating the affine
parameterization to the cosmological redshift. In Section~III we
present our main, model-independent argument. In Section~IV we
offer some remarks on the interpretation and possible extensions
of our result.

\medskip

\noindent \textit{II. A simple model.}
Consider a model in which the metric takes the form
\begin{equation}
ds^2=dt^2-a^2(t)d{\vec x}^{\,2}.
\label{eq:Flat}
\end{equation}
We will first examine the behavior of null geodesics, and then
timelike ones.

From the geodesic equation one finds that a null geodesic in the
metric~(\ref{eq:Flat}), with affine parameter $\lambda$, obeys
the relation
\begin{equation}
     d \lambda \propto a(t) \, d t .
     \label{eq:AP}
\end{equation}
Alternatively, we can understand this equation by considering a
physical wave propagating along the null geodesic.  In the short
wavelength limit the wave vector $k^\mu$ is tangential to the
geodesic, and is related to the affine parameterization of the
geodesic by $k^\mu \propto d x^\mu / d \lambda $.  This allows us
to write $d \lambda \propto d t / \omega$, where $\omega \equiv
k^0$ is the physical frequency as measured by a comoving
observer. In an expanding model the frequency is red-shifted as
$\omega \propto 1/a(t)$, so we recover the result
of Eq.~(\ref{eq:AP}).

From Eq.~(\ref{eq:AP}), one sees that if $a(t)$ decreases
sufficiently quickly in the past direction, then $\int a(t) \, d
t$ can be bounded and the maximum affine length must be finite.
To relate this possibility to the behavior of the Hubble
parameter $H$, we first normalize the affine parameter by
choosing $d \lambda = \left[a(t) / a(t_f)\right] d t$, so $d
\lambda / dt \equiv 1$ when $t=t_f$, where $t_f$ is some chosen
reference time. Using $H \equiv \dot a / a$, where a dot denotes
a derivative with respect to~$t$, we can multiply Eq.~(\ref{eq:AP})
by $H(\lambda)$ and then integrate from some
initial time $t_i$ to the reference time $t_f$:
\begin{equation}
\int_{\lambda(t_i)}^{\lambda(t_f)} \, H(\lambda) \, d \lambda =
     \int_{a(t_i)}^{a(t_f)} {d a \over a(t_f)} \le 1 ,
\label{eq:nulleq}
\end{equation}
where equality holds if $a(t_i)=0$.  Defining
$H_{\rm av}$
to be an average over the affine parameter,
\begin{equation}
H_{\rm av} \equiv {1 \over \lambda(t_f)-\lambda(t_i)}
\int_{t_i}^{t_f} \, H(\lambda) \, d \lambda \le {1 \over
\lambda(t_f)-\lambda(t_i)} \ ,
\label{eq:nullbound}
\end{equation}
we see that any backward-going null geodesic with $H_{\rm av} > 0$ must
have a finite affine length, i.e., is past-incomplete.

A similar argument can be made for timelike geodesics,
parameterized by the proper time $\tau$.  For a particle of mass
$m$, the four-momentum $P^\mu \equiv m \, d x^\mu / d \tau $, so
we can write $d \tau = (m/E) \, d t$, where $E \equiv P^0$ is the
energy of the particle as measured by a comoving observer.  If we
define the magnitude of the three-momentum $p$ by $p^2 \equiv -
g_{i j} \, P^i \, P^j$, where $i$ and $j$ are summed 1 to 3, then
$E = \sqrt{p^2 + m^2}$. For a comoving trajectory we have $P^i =
0$, and therefore $d \tau = d t$. For all others, $p \propto
1/a(t)$~\footnote{This follows from the spatial components of the
geodesic equation of motion for the particle: $d(a^2P^i)/d\tau=
\hbox{$-(m/2) \partial_i g_{\mu \nu} (d x^\mu / d \tau) (d x^\nu
/ d \tau) = 0 $, where $p^2 = a^2 \left(P^i\right)^2$}$.}, so we
can write $p(t) = [a(t_f)/a(t)] \, p_f$, where
$p_f$ denotes the value of the
three-momentum $p$ at the reference time $t_f$.
Combining all this, we find
\begin{eqnarray}
&&\int_{t_i}^{t_f} \, H(\tau) \, d \tau = \int_{a(t_i)}^{a(t_f)}
     { m \, da \over \sqrt{ m^2 \, a^2 + p_f^2 \, a^2 (t_f) }}
     \nonumber \\
&&\qquad \le \ln \left( {E_f + m \over p_f } \right) ={1 \over 2}
     \ln \left( {\gamma + 1 \over \gamma - 1} \right) \, ,
\label{eq:timeeq}
\end{eqnarray}
where the inequality becomes an equality when $a(t_i) = 0$.  Here
$E_f \equiv \sqrt{p_f^2+m^2}$ and $\gamma \equiv 1/\sqrt{1-v_{\rm
rel}^2}$, where $v_{\rm rel} \equiv p_f/E_f$ is the speed of the
geodesic relative to the comoving observers at time $t_f$. Since
the integral is bounded, the argument used for null geodesics can
be repeated, with the average taken over proper time.

\medskip

\noindent
\textit{III. The main argument.}
In this section we show that the inequalities of
Eqs.~(\ref{eq:nulleq}) and (\ref{eq:timeeq}) can be established
in arbitrary cosmological models, making no assumptions about
homogeneity, isotropy, or energy conditions.  To achieve this
generality, we need a definition of the Hubble parameter~$H$ that
applies to arbitrary models, and which reduces to the standard
one ($H=\dot a/a$) in simple models.

Consider a timelike or null geodesic $\cal O$ (``the observer'').
We assume that a congruence of timelike geodesics (``comoving
test particles'') has been defined
along~$\cal O$~\footnote{We do not require that this congruence be defined
throughout spacetime, just along $\cal O$. Away from $\cal O$ the
members of this congruence may cross or focus, but such behavior
does not affect our argument.},
and we will construct a definition for~$H$ that depends only on
the relative motion of the observer and test particles.

In order to motivate what we do, we first consider the case of
nonrelativistic velocities in Minkowski space.  Suppose that the
observer measures the velocities of the test particles as a
function of the time~$\tau$ on his own clock. At time~$\tau_1$
particle~1 passes with velocity $\vec u(\tau_1)$, and at time
$\tau_2 = \tau_1+\Delta\tau$ particle~2 passes with
velocity~$\vec u(\tau_2)$. What expansion rate could he infer
from these measurements? The separation vector between the
positions~${\vec r}_1$ and~${\vec r}_2$ of the two particles at
$\tau_2$ is $\Delta \vec r={\vec r}_1-{\vec r}_2=\vec u \, \Delta
\tau$, and its magnitude is $\Delta r\equiv|\Delta \vec r \,|$.
Their relative velocity is $\Delta \vec u={\vec u}_1- {\vec
u}_2=-(d\vec u/d \tau) \Delta \tau$. The Hubble expansion rate is
defined in terms of the rate of separation of these particles,
which in turn depends on the radial component of their relative
velocity, $\Delta u_r = \Delta \vec u \cdot \Delta \vec r/\Delta
r$. The inferred Hubble parameter $H$ is then
\begin{equation}
H \equiv {\Delta u_r \over \Delta r}
  = {\Delta \vec u \cdot \Delta \vec r \over |\Delta \vec r \,|^2}
  = - {\vec u \cdot ({d \vec u / d \tau}) \over
     | \vec u \,|^2} \, ,
\label{eq:HdefNR}
\end{equation}
which will equal
the standardly defined Hubble parameter for the case of a
homogeneous, isotropic universe. The expression for $H$ may be
simplified to $H = - d \, \ln v_{\rm rel} / d \tau$, where
$v_{\rm rel} = |\vec u\,|$. The fact that $H$ is the
total derivative of a function of $v_{\rm rel}$ implies
that the variation of $v_{\rm rel}$ is determined completely
by the local value of
$H$, even if the universe is inhomogeneous and anisotropic.

We can now generalize this idea to the case of arbitrary
velocities in curved spacetime. Let $v^\mu = d x^\mu / d \tau$ be
the four-velocity of the geodesic $\cal O$, where we take $\tau$
to be the proper time in the case of a timelike observer, or an
affine parameter in the case of a null observer.  Let
$u^\mu(\tau)$ denote the four-velocity of the comoving test
particle that passes the observer at time $\tau$. We define
$\gamma \equiv u_\nu v^\nu$, so in the timelike case $\gamma$ may
be viewed as the relative Lorentz factor
($1/\sqrt{1-v_{\text{rel}}^2}$) between $u^\mu$ and $v^\mu$. In
the null case, $\gamma = d t /d\tau$, where $t$ is the time as
measured by comoving observers, and $\tau$ is the affine
parameter of $\cal O$.

\begin{figure}
\includegraphics[scale=.8]{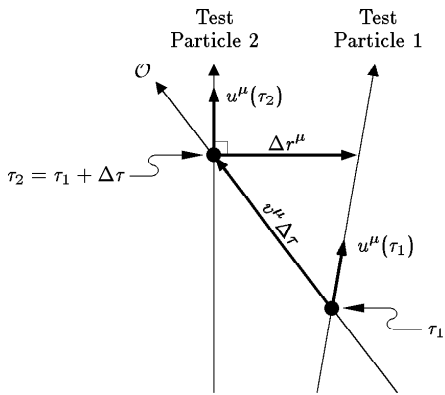}
\caption{\label{fig:one} The observer's worldline $\cal O$ and
two test particles.}
\end{figure}

Consider observations made by ${\cal O}$ at times $\tau_1$ and
$\tau_2 = \tau_1 + \Delta \tau$, as shown in Fig.~\ref{fig:one},
where $\Delta \tau$ is infinitesimal. Let $\Delta r^\mu$ be a
vector that joins the worldlines of the two test particles at
equal times in their own rest frame.  Such a vector is
perpendicular to the worldlines, and can be constructed by
projecting the vector $-v^\mu \Delta \tau$ to be perpendicular to
$u^\mu$: $\Delta r^\mu = - v^\mu \Delta \tau + \gamma u^\mu
\Delta \tau$. In the rest frame of the observer this answer
reduces for small velocities to its nonrelativistic counterpart,
$\Delta r^\mu = (0, u^i \, \Delta \tau)$. This is a spacelike
vector of length $\Delta r \equiv |\Delta r^\mu|=\sqrt{\gamma^2 -
\kappa} \, \Delta \tau$, where $\kappa \equiv v^\mu v_\mu$ is
equal to $1$ for the timelike case and $0$ for the null case. The
separation velocity will be $\Delta u^\mu = - (Du^\mu / d \tau)
\, \Delta \tau$, where $D/d \tau$ is the covariant derivative
along $\cal O$. The covariant derivative allows us to compare via
parallel transport vectors defined at two different points along
${\cal O}$, and can be justified by considering the problem in
the free-falling frame, for which the affine connection vanishes
on ${\cal O}$ at $\tau=\tau_1$. The radial component of this
velocity will be $\Delta u_r = - \left(\Delta u^\mu \Delta r_\mu
\right) / \Delta r$, where the sign arises from the Lorentz
metric. We define the Hubble parameter as
\footnote{Our definition of $H$ in Eq.~(8) can also be expressed as $H =
-u^\mu_{;\nu}n_\mu n^\nu$, where $n^\mu$ is a unit vector in the direction
of $\Delta r^\mu$. This can be shown by substituting\break $v^\mu=\gamma
u^\mu - \sqrt{\gamma^2-\kappa}n^\mu$ into~(8) and using the relations
$u_{\mu;\nu}u^\mu=u_{\mu;\nu}u^\nu=0$. This expression shows that $H$ depends
on the direction but not the speed of $\cal O$, as seen in the rest frame of
the comoving test particles.  Furthermore, the frequently used definition
$\tilde H = (1/3)u^\mu_{;\mu}$ is obtained from ours by averaging over all
directions.}
\begin{equation}
H \equiv {\Delta u_r \over \Delta r} = {- v_\mu (Du^\mu/ d \tau)
     \over {\gamma^2 - \kappa}} \, .
\end{equation}

Since $\cal O$ is a geodesic, we have $(D v^\mu / d \tau) = 0$,
and therefore
\begin{equation}
H= {- d\gamma/d \tau \over {\gamma^2 -
     \kappa}}={d\hphantom{\tau}\over d \tau} F\bigl(\gamma(\tau)\bigr),
\end{equation}
where
\begin{equation}
F(\gamma)=\cases{\gamma^{-1} & \text{null observer ($\kappa=0$)}\cr
        &\cr
        \displaystyle {1\over 2}
        \ln {\gamma + 1\over \gamma -1} &
        \text{timelike observer ($\kappa=1$)}\cr}
\end{equation}
As in Section~II, we now integrate~$H$ along ${\cal O}$ from some
initial $\tau_i$ to some chosen $\tau_f$:
\begin{equation}
\int_{\tau_i}^{\tau_f} \! H \, d\tau = F(\gamma_f)-F(\gamma_i) \le
     F(\gamma_f) \, .
\label{eq:bound}
\end{equation}
In the null case $F(\gamma_f) = \gamma_f^{-1}$, which is equal to
the value of $d \tau / d t$ at $t_f$, normalized in Section~II to
unity.

Eq.~(\ref{eq:bound}) therefore reproduces exactly the results of
Eqs.~(\ref{eq:nulleq}) and~(\ref{eq:timeeq}), but in a much more
general context.  Again we see that if $H_{\rm av} > 0$ along any
null or non-comoving timelike geodesic, then the geodesic is
necessarily past-incomplete.

\medskip

\noindent
\textit{IV. Discussion.}
Our argument shows that null and timelike geodesics are, in
general, past-incomplete in inflationary models, whether or not
energy conditions hold, provided only that the averaged expansion
condition $H_{\rm av} > 0$ holds along these past-directed
geodesics. This is a stronger conclusion than the one arrived at
in previous work~\cite{BVOne} in that we have shown under
reasonable assumptions that almost all causal geodesics, when
extended to the past of an arbitrary point, reach the boundary of
the inflating region of spacetime in a finite proper time (finite
affine length, in the null case).

What can lie beyond this boundary? Several possibilities have
been discussed, one being that the boundary of the inflating
region corresponds to the beginning of the Universe
in a quantum nucleation event~\cite{VilenkinWV}. The boundary is
then a closed spacelike hypersurface which can be determined from
the appropriate instanton.

Whatever the possibilities for the boundary, it is clear that
unless the averaged expansion condition can somehow be avoided
for all past-directed geodesics, inflation alone is not
sufficient to provide a complete description of the Universe, and
some new physics is necessary in order to determine the correct
conditions at the boundary~\footnote{Aguirre and Gratton~\cite{Aguirre-Gratton}
have proposed a model in which this new physics is in fact
also inflation, but inflation in the time-reversed sense.}.
This is the chief result of our
paper. The result depends on just one assumption: the Hubble
parameter $H$ has a positive value when averaged over the affine
parameter of a past-directed null or noncomoving timelike
geodesic.

The class of cosmologies satisfying this assumption is not
limited to inflating universes.  Of particular interest is the
recycling scenario~\cite{GaVi}, in which each comoving region
goes through a succession of inflationary and thermalized epochs.
Since this scenario requires a positive true vacuum energy
$\rho_{v}$, the expansion rate will be bounded by $H_{\min}=
\sqrt{8\pi G \rho_{v}/3}$ for locally flat or open equal-time
slicings, and the conditions of our theorem may be satisfied. One
must look carefully, however, at the possibility of
discontinuities where the inflationary and thermalized regions
meet. This issue requires further analysis.

Our argument can be straightforwardly extended to cosmology in
higher dimensions. For example, in the model of
Ref.~\cite{Bucher} brane worlds are created in collisions of
bubbles nucleating in an inflating higher-dimensional bulk
spacetime. Our analysis implies that the inflating bulk cannot be
past-complete.

We finally comment on the cyclic universe model~\cite{SteinTur}
in which a bulk of 4 spatial dimensions is sandwiched between two
3-dimensional branes.  The effective $(3+1)$-dimensional geometry
describes a periodically expanding and recollapsing universe,
with curvature singularities separating each cycle.  The internal
brane spacetimes, however, are nonsingular, and this is the basis
for the claim~\cite{SteinTur} that the cyclic scenario does not
require any initial conditions.  We disagree with this claim.

In some versions of the cyclic model the brane spacetimes are
everywhere expanding, so our theorem immediately implies the
existence of a past boundary at which boundary conditions must be
imposed.  In other versions, there are brief periods of
contraction, but the net result of each cycle is an expansion. 
For null geodesics each cycle is identical to the others, except
for the overall normalization of the affine parameter.  Thus, as
long as $H_{\rm av} > 0$ for a null geodesic when averaged over
one cycle, then $H_{\rm av} > 0$ for any number of cycles, and
our theorem would imply that the geodesic is incomplete.

\begin{acknowledgments}
We are grateful to Jaume Garriga, Gary Gibbons, Andrei Linde, 
and an anonymous
referee of this paper for useful comments.  Two of
us (AB and AHG) thank the Institute of Cosmology at
Tufts University for its hospitality. Partial financial support
was provided for AB by the Research Awards Committee of
Southampton College, for AHG by the U.S. Department of Energy
(D.O.E.) under cooperative research agreement
\#DF-FC02-94ER40818, and for AV by the National Science
Foundation.
\end{acknowledgments}

\end{document}